\pdfoutput=1
\documentclass[aps,pre,showpacs,twocolumn]{revtex4-1}
\usepackage{amssymb}
\usepackage{amsmath}
\usepackage{graphicx}
\usepackage{color}
\usepackage{soul}

\begin{document}

\title{Quantfying rich patterns in agglomeration of floating beads}

\author{Ceyda Sanl{\i}}
\email{cedaysan@gmail.com}
\altaffiliation{Present address: CompleXity Networks, naXys, University of Namur, 5000 Namur, Belgium}
\author{Detlef Lohse}
\email{d.lohse@utwente.nl}
\author{Devaraj van der Meer}
\email{d.vandermeer@utwente.nl}
\affiliation{Physics of Fluids group, MESA+ Institute for Nanotechnology, J. M. Burgers Centre for Fluid Dynamics, University of Twente, P.O. Box 217, 7500 AE Enschede, The Netherlands}

\date{\today}

\begin{abstract}
Macroscopic spherical particles spontaneously form rich patterns on a standing Faraday wave. These patterns are found to follow a very systematic trend depending on the floater concentration $\phi$: The same floaters that accumulate at amplitude maxima (antinodes) of the wave at low $\phi$, surprisingly move towards the nodal lines when $\phi$ is beyond a certain value. In more detail, circular irregularly packed antinode clusters at low $\phi$ give way to loosely packed filamentary structures at intermediate $\phi$, and are then followed by densely packed grid-shaped node clusters at high $\phi$. Here, we successfully characterize the morphology of these rich patterns using a metric analysis, i.e., the Minkowski functionals. We modify the Minkowski functionals such that we are able to measure the physical quantities of the clusters such as area, perimeter, and aspect ratio. 
\end{abstract}

\pacs{$47.54.-$r, $05.65.+$b}


\maketitle


\section{Introduction}\label{intro} 

Mountains, plateaus, valleys, rivers, lakes, and oceans decorate our earth at different scales. Not only how this variety of shapes presents itself but also how the shapes are connected to each other are significant questions addressed by geologists. However, why we live in such a decorated world is a matter hasn't been integrated into fundamental science. Are there universal laws that would explain all these rich formations from large-scale galaxies to small-scale cell tissues?

Physical, biological, and economical systems show emergence of collective behavior~\cite{Helbing}. Spins align themselves even without any external field and exhibit a spontaneous magnetization~\cite{Chandler}. A swirling motion in vertically shaken dry granular media~\cite{AransonVolfsonTsimring} and stratification patterns induced by granular avalanche in a rotating drum~\cite{GrayHutter} are observed (see Ref.~\cite{AransonTsimring} for an extensive review).

Biological cells form patterns spontaneously~\cite{KochMeinhardt, Karsenti} and swarms of birds move collectively in a randomly chosen direction, and so reach the emergence of ordered states~\cite{AldanaDossettiHuepeKenkreLarralde}. In economical systems, a restricted amount of resources in a group of people force them to form small groups connected by a special network~\cite{HSimon}.

In this paper, we study the pattern formation of macrospheres floating on the surface of a standing Faraday wave. When plastic spheres are exposed to a periodic oscillation at a water-air interface they exhibit well-organized clusters and, beyond a transitional value of the floater concentration $\phi$, emergence of collective behavior dominates the system~\cite{antinodenodePRE, CSanliThesis}:  The same spheres that cluster at the wave maxima (antinodes) at low $\phi$, cluster at the wave minima (nodes) at high $\phi$. Single hydrophilic heavy spheres on a standing Faraday wave experience a drift towards the antinodes~\cite{Falkovich, FalkovichPRL, FalkovichEurPhys}. However, in the presence of the attractive gravity-induced capillary interaction~\cite{Capillaryforce, MichealBerhanu, PlanchetteLorenceauBiance, PotyLumayVandewalle}, not only the position of the clusters but also their dynamics is completely different: The antinode clusters breath but the node clusters do not. In our previous work~\cite{antinodenodePRE, CSanliThesis}, by integrating the breathing phenomenon into the drift, we have qualitatively and quantitatively explained both experimentally observed limits, i.e., the antinode clusters at low $\phi$ and the node clusters at high $\phi$. However, in a series of controlled experiments at intermediate $\phi$, we observe morphologically rich patterns that haven't been addressed in the previous work: Circular irregularly packed antinode clusters at low $\phi$ are replaced by loosely packed filamentary structures at intermediate $\phi$, and then followed by densely packed grid-shaped node clusters at high $\phi$. Here, we now calculate and successfully interpret the morphological details of these patterns.

Several types of metric analysis has been applied to characterize heterogeneity of colloids~\cite{LiuNagelSaarloosWyart}, collective movements of macroscopic particles~\cite{LechenaultDauchotBiroliBouchaud} and animals~\cite{GrossmanAransonJacob}, and human mobility patterns~\cite{GonzalezHidalgoBarabasi}. Most of them focus on constructing either individual spatiotemporal trajectories or a correlation of particle positions in two (or four) points in space (and/or time). In this paper, we suggest an alternative approach, inspired by the Minkowski functional point pattern~\cite{MeckeStoyanSensitivityofMethod, MeckeBuchertWagner, MeckeReview}, and show that we are able to quantify the physical characteristics of the patterns such as the cluster size, perimeter, and aspect ratio.

\section{Experiment}\label{exp} 

\subsection{Setup}\label{Sc:Setup}

\begin{figure}[h!]
\begin{center}
  \includegraphics[width=8.2 cm]{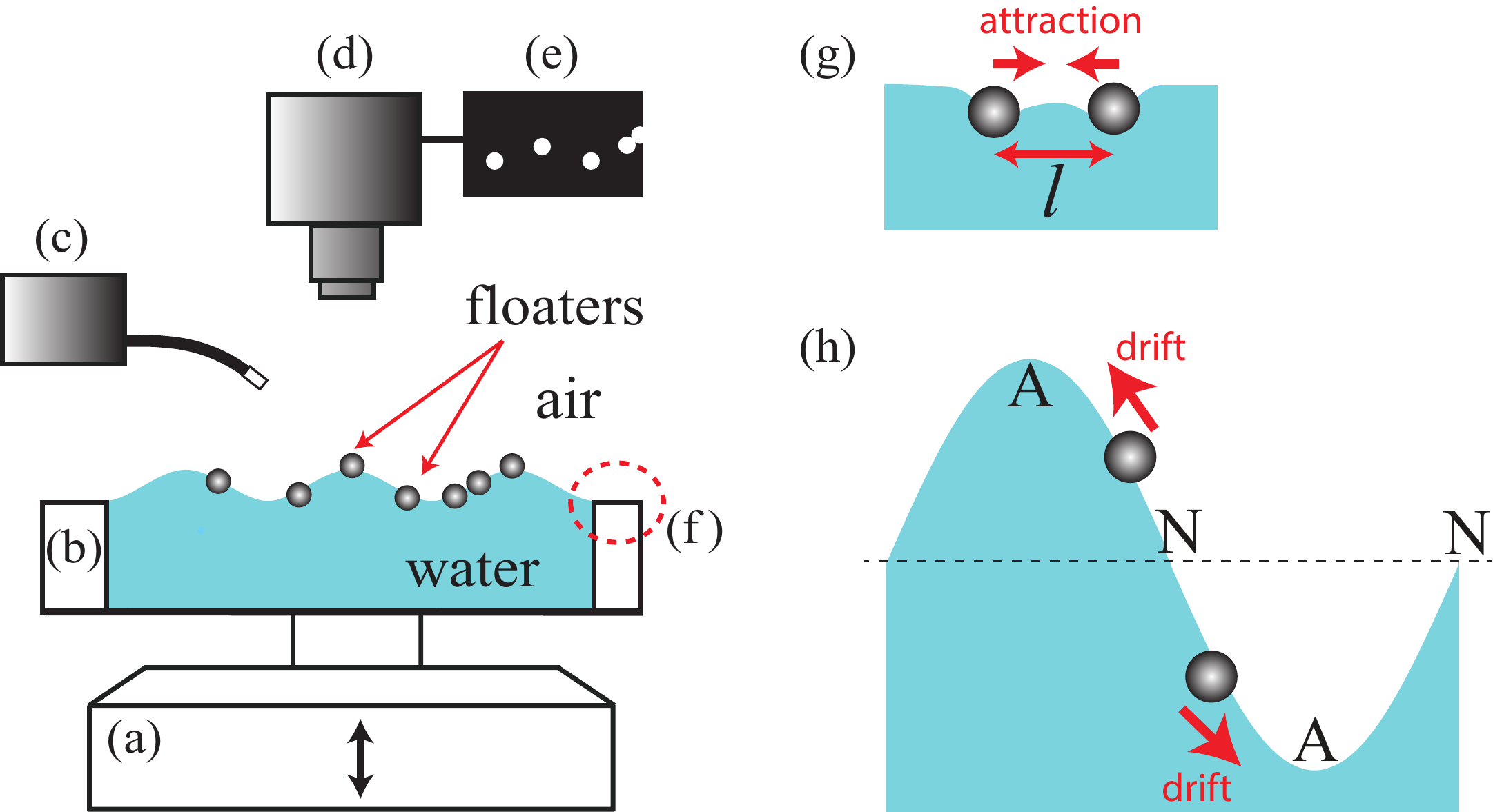}
\end{center}
\vspace{-6mm} \caption{\label{Fig1_setup} \small (Color online). Experimental setup: (a) shaker, (b) glass container, 81$\times$45$\times$10 mm$^3$, (c) Schott fiber light source, (d) Photron Fastcam SA.1, (e) an illustration of a camera image, (f) pinned brim-full boundary condition, (g) the surface deformation around our hydrophilic heavy floaters causes an attractive force, (h) the direction of the period-averaged drift of a single floater, where A and N represent the antinode and the node, respectively.}
\end{figure}

The experimental setup is the same as we have used in our previous work~\cite{antinodenodePRE} and illustrated in Fig. \ref{Fig1_setup}. A container, made from transparent hydrophilic glass with 10 mm height and 81$\times$45 mm$^2$ rectangular cross section is attached to a shaker. The container is completely filled with purified water (Millipore water with a resistivity $>$ 18 M$\Omega\cdot$cm) such that the water level is perfectly matched with the container edge [Fig. \ref{Fig1_setup}(f)]. Using this so-called brim-full boundary condition~\cite{Douady}, a static surface inclination induced by the boundary is avoided~\cite{Falkovich, FalkovichPRL, FalkovichEurPhys}. Spherical polystyrene floaters (contact angle $74^\circ$ and density 1050 kg/m$^3$) with average radius $R$ around 0.31 mm and a polydispersity of approximately $14\%$ are carefully distributed over the water surface to make a monolayer. To avoid surfactant effects, we clean both the container and the floaters by performing the cleaning protocol described in Ref.~\cite{cleaningprotocol}. (See Ref.~\cite{antinodenodePRE} for further information on the determination of the contact angle of the floater.)

A standing Faraday wave is generated using a shaker providing a vertical sinusoidal oscillation with amplitude $a_0$ and frequency $f_0$. We determine $f_0$ such that we produce a rectangular wave pattern with a wavelength in the range of $17$ to $24$ mm corresponding to frequencies ranging from $37$ to $42$ Hz (note that the standing Faraday wave frequency is equal to $f_0/2$).  
Adding floaters to the surface, we need to slightly adjust both $a_0$ and $f_0$ to obtain a well defined rectangular pattern. (More details of the procedure for creating a standing Faraday wave in the presence of floaters can be found in Ref.~\cite{antinodenodePRE}.)
A continuous white fiber-light source (Schott) is used to illuminate the floaters from the side as shown in Fig. \ref{Fig1_setup}(c). The two-dimensional floater positions are recorded with a high-speed camera (Photron Fastcam SA.1) at 500 frames per second. 
Each image is 546$\times$1030 pixels (38$\times$72 mm$^2$), which covers around $75\%$ of the total cross section area of the container. 
The vertical depth of field is taken to be large enough to capture the maximum vertical displacement ($2.5\pm0.1$ mm) of the floaters.

In the period-averaged context, there are two mechanisms that drive the floaters on the standing Faraday wave. The first one is the attractive capillary interaction~\cite{Capillaryforce, MichealBerhanu, PlanchetteLorenceauBiance, PotyLumayVandewalle} due to the surface deformation around the floaters [Fig. \ref{Fig1_setup}(g)], which is significant  
when the distance between the floaters $l$ is smaller than the capillary length $l_c=(\sigma/\rho g)^{1/2}$. Here, $\sigma$ is the surface tension coefficient of the interface, $\rho$ the liquid density, and $g$ the acceleration of gravity. (For an air-water interface at $20\,^\circ$C, $l_c=2.7$ mm.) The second is due to 
the standing Faraday wave, which causes  
a time-averaged drift of the floaters towards the antinodes [Fig.~\ref{Fig1_setup}(h)], which is observed and described in~\cite{antinodenodePRE, Falkovich, FalkovichEurPhys}. 

\subsection{Floater patterns}\label{Sc:FloaterPatterns}

\begin{figure}[h!]
\begin{flushleft}
  \includegraphics[width=8.5 cm]{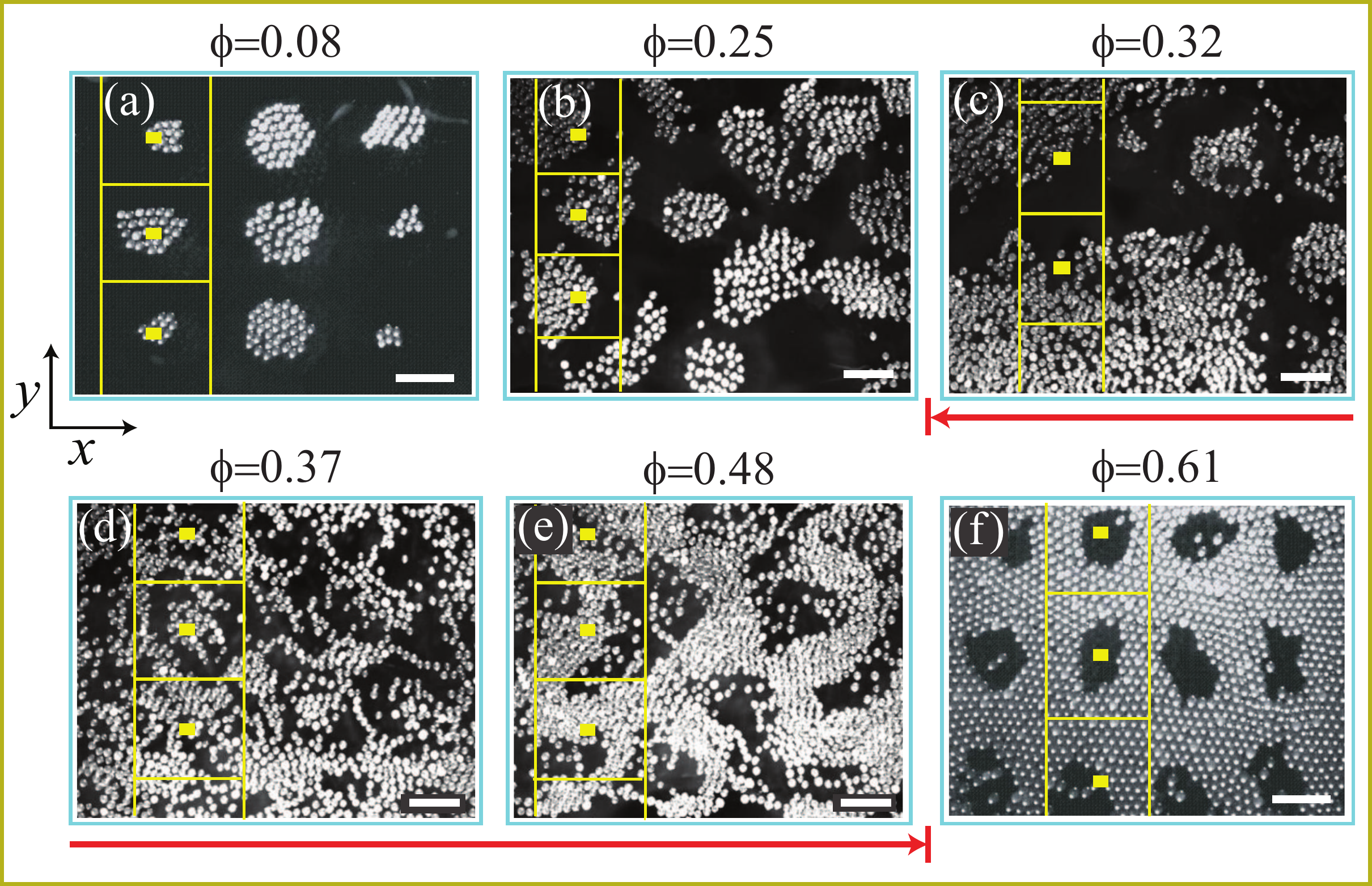}
\end{flushleft}
\vspace{-6mm} \caption{\small (Color online) Floater patterns on a standing Faraday wave with a rectangular wave pattern for various floater concentrations $\phi$. To indicate the corresponding wave structures, in a small subset of the field of view the antinodes and the nodes are presented by small yellow rectangles and yellow lines, respectively. The images which lie in (or nearby) the transition region from the antinode to the node clusters are marked by the red line. The morphological differences in the clusters are clear: At low $\phi$, circular, irregularly packed antinode clusters are observed (a, b). At intermediate $\phi$, when we enter the transition region, clusters with a large heterogeneity in space evolve (c). Further into the transitional region, loosely packed filamentary structures occur (d), which when $\phi$ increases even more evolve into densely clustered regions (e) from which finally densely packed grid-shaped node clusters form (f). The white bars indicate a length scale of 5 mm.}\label{Fig2_FloaterPatterns}
\end{figure}

The experimentally observed floater patterns on a standing Faraday wave are shown in Fig.~\ref{Fig2_FloaterPatterns}. The snapshots are taken from the top at a phase in which the wave elevation is nearly zero. For all cases, the Faraday wave pattern itself is rectangular with wavelengths $\lambda_x\approx\lambda_y\approx20 R$ for the $x$ and $y$-direction, respectively. The patterns result from the competition of the attractive capillary and the drift forces described in Section \ref{Sc:Setup}. Therefore, on physical grounds, our observed patterns presented in Fig.~\ref{Fig2_FloaterPatterns}(a)-\ref{Fig2_FloaterPatterns}(f) have different origins compared to the Faraday wave patterns occurring on the free surface without floaters~\cite{Douady, GluckmanArnoldGollubchaoticwavepatterns, BinksvandeWater, ZhangVinals, KudrolliAbrahamGollub}.

Clearly, the route from the antinode clusters at low $\phi$ [Fig.~\ref{Fig2_FloaterPatterns}(a)] to the node clusters at high $\phi$ [Fig.~\ref{Fig2_FloaterPatterns}(f)] includes many different stages of varying particle mobility: We observe mobile antinode clusters [Fig.~\ref{Fig2_FloaterPatterns}(b)], large, loosely packed clusters [Fig.~\ref{Fig2_FloaterPatterns}(c)], loosely packed filamentary clusters [Fig.~\ref{Fig2_FloaterPatterns}(d)], and densely clustered regions [Fig.~\ref{Fig2_FloaterPatterns}(e)]. It is clear that, the quantitative analysis of such variety of structures and packings poses a considerable challenge and will be addressed the next Sections. 


\section{Complex pattern analysis}\label{Sc:Complexpatternanalysis} 

\subsection{Minkowski point pattern approach}\label{Sc:Minkowski point pattern approach}

\begin{figure}[h!]
\begin{center}
  \includegraphics[width=9.0 cm]{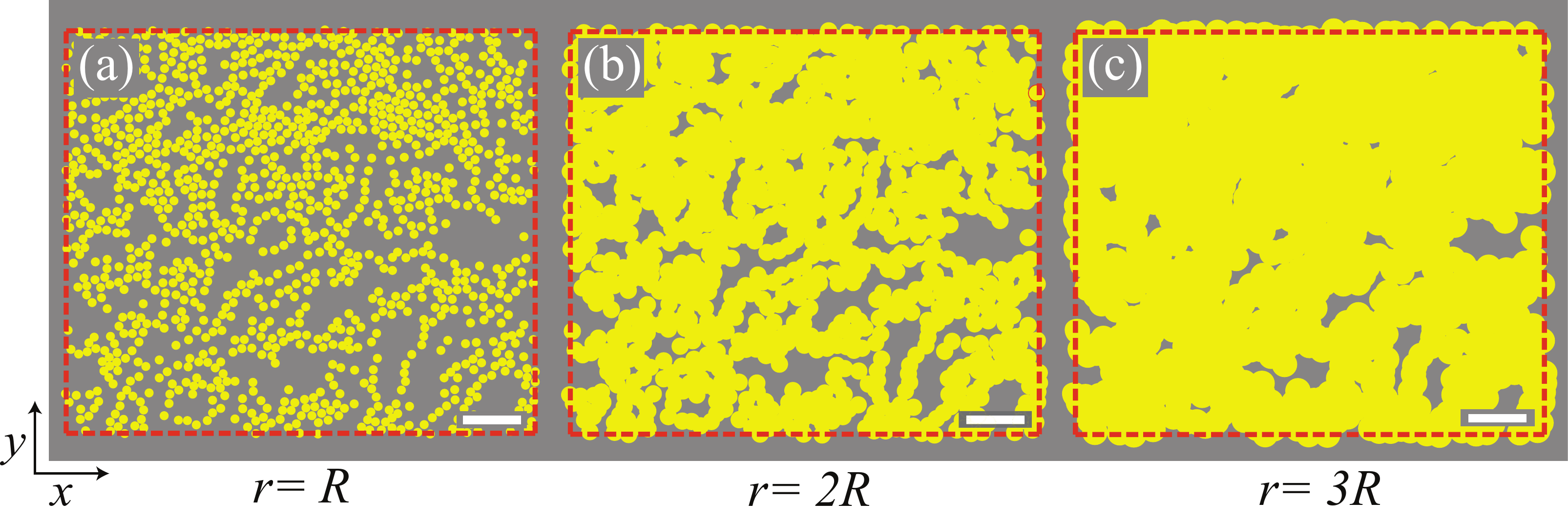}
\end{center}
\vspace{-6mm} \caption{\small The Minkowski point pattern approach is demonstrated using the particle pattern at the intermediate floater concentration $\phi=0.37$ [see Fig.~\ref{Fig2_FloaterPatterns}(d) for the corresponding experimental image]. Here, we show the morphological deformation of the particle pattern when the Minkowski radius $r$ increases from the particle radius $R$ (a), which provides an image identical to the experimental one, to twice the radius $2R$ (b), and finally to three times the radius $r=3R$ (c). The dashed (red) lines are the boundaries of the corresponding experimental image. The white bars indicate a length scale of 5 mm.}\label{Fig3_MinkowskiPointPattern}
\end{figure}

A system of particles can be viewed as a collection of points defined by the positions of the particle centers. To study the morphological properties of the set of points (such as clustering), discs of radius $r$ are drawn around each particle center and then $r$ is varied. In Fig.~\ref{Fig3_MinkowskiPointPattern}, we demonstrate this process for an intermediate floater concentration ($\phi=0.37$, in the transition region) as an example of a pattern constructed with discs of increasing radius $r$: From the particle radius $R$ to $3R$, the resultant pattern changes due to the merging of isolated particles into groups (of particles), as shown in Fig.~\ref{Fig3_MinkowskiPointPattern}(a) to Fig.~\ref{Fig3_MinkowskiPointPattern}(c). In the first figure, all particles are separate, whereas in the second almost all become connected. The large number of holes disappears when $r$ further increases as illustrated by the last figure.

The morphology of the point patterns can be quantified by the Minkowski functionals, which are introduced in the next Section.

\subsection{Minkowski functionals}\label{Sc:Minkowskifunctionals}

The Minkowski functionals are a set of functionals providing information of geometry, shape (i.e. curvature), and connectivity of any pattern of interest. They have been successfully employed to characterize large-scale structures such as galaxy samples~\cite{MeckeBuchertWagner}, turbulent and regular patterns in chemical reaction-diffusion systems~\cite{Mecke}, spinodal decomposition structures in a simulation~\cite{MeckeSofonea} and also in an experiment of a colloid-polymer mixture~\cite{marchingalgorithm}, particle and bubble distributions in turbulent flow~\cite{CalzavariniKerscherLohseToschi}, and aggregations of colloids at fluid interfaces~\cite{BleibelDominguezOettelDietrich}. Detailed extensive reviews can be found in Refs.~\cite{MeckeReview, MichielsenRaedtReview}, and the sensitivity of the method for some data sets in comparison with the other methods are discussed in Ref.~\cite{MeckeStoyanSensitivityofMethod}.

The Minkowski functionals $\widetilde{M}_0(D), \widetilde{M}_1(D), \widetilde{M}_2(D),\,\ldots\, , \widetilde{M}_{d}(D)$ are functionals of integral geometry, which are additive, motion-invariant, and continuous on a given domain $D$ in $d-$dimensional Euclidean space~\cite{MeckeSofonea}. Mathematically, the $\widetilde{M}_\nu(D)$ are defined as~\cite{marchingalgorithm}
\begin{widetext}
\begin{eqnarray}\label{Eq:Minkowskigeneraldefinition}
    \widetilde{M}_0(D)&=&\int\limits_D\,d^d\vec{r}, \\ \nonumber
    \widetilde{M}_\nu(D)&=&\frac{\omega_{d-\nu}}{\omega_\nu\omega_d}\frac{(\nu-1)!(d-\nu)!}{d!}\int\limits_{\partial D}\sum\limits_{\{i_1,\,\ldots\, , i_{\nu-1}\}}\frac{1}{R_{i_1}\ldots R_{i_{\nu-1}}}\,d^{d-1}\vec{r},\,\mbox{when }\nu\geq1,
\end{eqnarray}
\end{widetext}
where $\omega_d=\pi^{d/2}/\Gamma(1+d/2)$ is the unit volume in $d$ dimensions, $\partial D$ is the boundary of the domain $D$, and $R_i$ are the principal radii of curvature with $i=1, \ldots , d-1$. Here, the summation is performed over the multiplication of all permutations $i_\nu$ of $d-1$ curvatures.

We will now compute the Minkowski functionals for domains $D$ consisting of the patterns introduced in Fig.~\ref{Fig3_MinkowskiPointPattern}. In two dimensions ($d=2,\,\nu\leq2$), using Eq.~\ref{Eq:Minkowskigeneraldefinition}, there are three Minkowski functionals~\cite{marchingalgorithm}
\begin{eqnarray}\label{Eq:2DMinkowskifunctionals}
  \widetilde{M}_0(D)&=&\int\limits_D\,d^2\vec{r}, \\ \nonumber
  \widetilde{M}_1(D)&=&\frac{1}{2\pi}\,\int\limits_{\partial D}\,d\vec{r}, \\ \nonumber
  \widetilde{M}_2(D)&=&\frac{1}{2\pi^2}\int\limits_{\partial D}\,\frac{1}{R}\,d\vec{r},
\end{eqnarray}
where $d^2\vec{r}$ and $d\vec{r}$ are the Lebesgue measures in two and one-dimensional Euclidean spaces, respectively. Here, $M_0(D)$ is the area of $D$, $M_1(D)$ is proportional to the perimeter of $D$, and $M_2(D)$ is proportional to the measure of the connectivity, called Euler characteristic (commonly labeled by $\chi$). In this study, instead of the exact values of the functionals, we are interested in their actual physical meanings. Therefore, as is also done in, e.g., Ref.~\cite{BleibelDominguezOettelDietrich}, we slightly redefine the prefactors in Eq.~\ref{Eq:2DMinkowskifunctionals}
\begin{eqnarray}\label{Eq:2DPhysicalMinkowskifunctionals}
 M_0(D)&=&\mbox{area of}\,D,\\ \nonumber
 M_1(D)&=&\mbox{perimeter of}\,D \,\Big(=2\pi\, \widetilde{M}_1(D)\Big), \\ \nonumber
 M_2(D)&=&\mbox{Euler characteristic of}\,D \,\Big(=\pi\, \widetilde{M}_2(D)\Big).
\end{eqnarray}

The Euler characteristic may appear to be a complicated quantity, but we point out that for a circular cluster with radius $R_\circ$ with the definition in Eq.~\ref{Eq:2DPhysicalMinkowskifunctionals} we simply have
\begin{equation}\label{Eq:2DEulercharacteristic_circulardomain}
M_2(D)=\frac{1}{2\pi}\int\limits_{0}^{2\pi}\,\frac{1}{R_\circ}\,R_\circ\,d\varphi=1,
\end{equation}
whereas for a circular hole of radius $R_\circ$, we obtain
\begin{equation}\label{Eq:2DEulercharacteristic_circularhole}
M_2(D)=\frac{1}{2\pi}\int\limits_{0}^{2\pi}\,\left(-\frac{1}{R_\circ}\right)\,R_\circ\,d\varphi=-1.
\end{equation}
The above results can be shown to hold for non-circular clusters and holes too, such that the Euler characteristic is simply found to be equal to the number of clusters minus the number of holes:
\begin{equation}\label{Eq:EulerCharacteristics}
    M_2(D)=\mbox{$\#$}_{\mbox{clusters}}-\mbox{$\#$}_{\mbox{holes}}.
\end{equation}
Therefore, if there are only distinct clusters, the Euler characteristic $M_2(D)>0$, whereas if there are only distinct holes $M_2(D)<0$. (In this paper, the connectivity or the Euler characteristic is called $M_2$ in contrast to the common notation $\chi$ used in the literature to keep the notation consistent with the other Minkowski functionals.)

We now introduce some further quantities that help us in the physical interpretation of the Minkowski functionals. To this end, $M_\nu(D)$ can be decomposed to two parts, one for the clusters and the other for the holes such that
\begin{itemize}\label{Eq:2DMinkowskifunctionals_clusters&holes}
\itemsep -1pt
  \item$[M_0(D)]_{\mbox{\footnotesize clusters}}$=total area of $D$, \hfill\vspace{0.5mm}\\
      \small(i.e., including the area occupied by the holes of $D$), \hfill \vspace{0.5mm}\\
  \item$[M_1(D)]_{\mbox{\footnotesize  clusters}}$=total external perimeter of $D$,\hfill\vspace{0.5mm}\\
      \small(i.e., disregarding the perimeter occupied by the holes of $D$), \hfill \vspace{1.5mm}\\
   \item$[M_0(D)]_{\mbox{\footnotesize holes}}$=total area of holes in $D$, \hfill\vspace{0.5mm}\\
       \small(i.e., total area occupied by the holes of $D$), \hfill \vspace{0.5mm}\\
  \item$[M_1(D)]_{\mbox{\footnotesize holes}}$=total perimeter of holes in $D$, \hfill\vspace{0.5mm}\\
      \small(i.e., total perimeter of the holes in $D$). \hfill
\end{itemize}
Furthermore, $M_2(D)$ is already in its decomposed form [Eq.~\ref{Eq:EulerCharacteristics}]. If we connect these decomposed quantities with the actual $M_0(D)$ and $M_1(D)$, we have
\begin{eqnarray}\label{Eq:2DPhysicalMinkowskifunctionalsconnecteddisconnectedgroups}
  M_0(D)&=&[M_0(D)]_{\mbox{\footnotesize clusters}}-[M_0(D)]_{\mbox{\footnotesize holes}}, \\ \nonumber
  M_1(D)&=&[M_1(D)]_{\mbox{\footnotesize clusters}}+[M_1(D)]_{\mbox{\footnotesize holes}}.
\end{eqnarray}
Both the clusters and the holes can be visualized for an arbitrary pattern. 

\subsection{Experimental results of the Minkowski functionals}\label{Sc:Experimentalresults}

The three Minkowski functionals in two dimensions are given by Eq.~\ref{Eq:2DPhysicalMinkowskifunctionals}. In this case, since the corresponding domain $D$ varies by increasing $r$, the functionals are called $M_\nu(r)$. Here, we also use their dimensionless forms
\begin{eqnarray}\label{Eq:2DPhysicalMinkowskifunctionalsdimensionlessforms}
  M^*_0(r)&=&\frac{M_0(r)}{N\,\pi r^2}, \\ \nonumber
  M^*_1(r)&=&\frac{M_1(r)}{N\,2\pi r}, \\ \nonumber
  M^*_2(r)&=&\frac{M_2(r)}{N},
\end{eqnarray}
where $N$ is the total number of floaters. With this non-dimensionalization we compare the area of the pattern $M_0(r)$ with the area of just as many non-overlapping discs of the same size, and similarly for $M_1(r)$ we divide the perimeter of the pattern with the sum of the perimeters of $N$ non-overlapping discs. Finally, $M_2(r)$ is non-dimensionalized by the Euler characteristic of $N$ non-overlapping spheres, which is $N$.

The reason behind this particular non-dimensionalization is that $M_0(r)$ and $M_1(r)$ trivially increase with the radius $r$ as the surface area $(\pi r^2)$ and perimeter $(2\pi r)$ of a disc of radius $r$. Non-dimensionalizing the Minkowski functionals as proposed in Eq.~\ref{Eq:2DPhysicalMinkowskifunctionalsdimensionlessforms} compensates for this trivial scaling of the Minkowski functionals. In addition, the extra factor $N$ by which the functionals are divided brings about that in the limit $r\ll R$, where we are dealing with $N$ separate discs, all three Minkowski functionals are equal to one: $M^*_0(r)=M^*_1(r)=M^*_2(r)=1$.

Note that in all experimental analysis, we apply the Minkowski point pattern approach by disregarding the experimental image boundaries such that, for each iteration of the disc radius $r$, an unbounded two-dimensional resultant pattern is considered as was shown in Fig.~\ref{Fig3_MinkowskiPointPattern}. 

\begin{figure*}[ht!]
\begin{center}
  \includegraphics[width=14.0 cm]{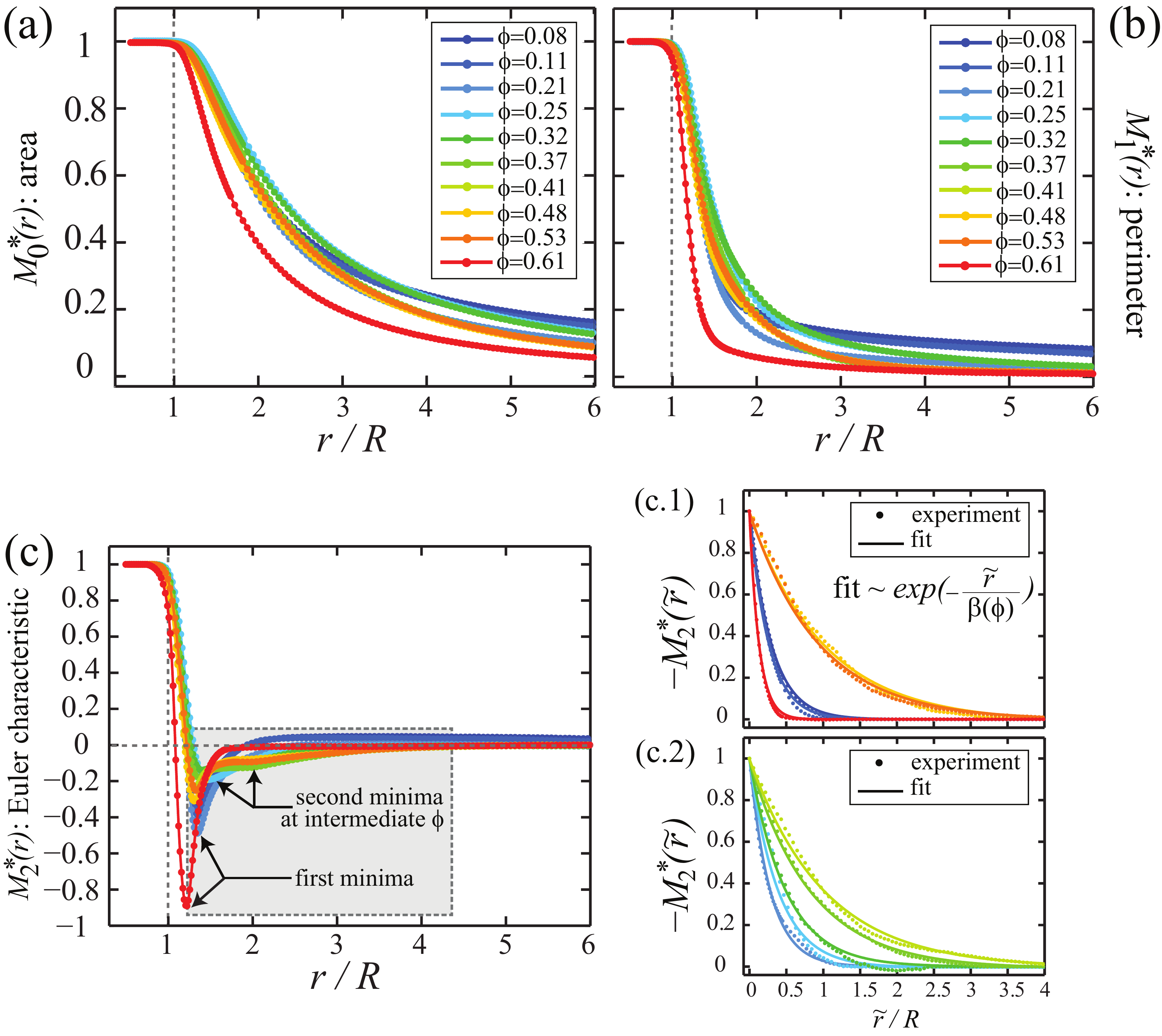}
\end{center}
\vspace{-6mm} \caption{\footnotesize The three dimensionless Minkowski functionals of the experimental data from antinode clusters at low $\phi$ (blue) to node clusters at high $\phi$ (red) as a function of $r$ normalized by the floater radius $R$. The normalized area $M^*_0(r)$ (a), the normalized perimeter $M^*_1(r)$ (b), and the Euler characteristic $M^*_2(r)$ (c) are presented. The decay in the curves when $r>R$ indicate an increasing amount of overlapping regions due to merging of the disks. Smoother decays at intermediate $\phi$ (relative to that at low and high $\phi$) suggest that more heterogenous and filamentary structures [Fig.~\ref{Fig2_FloaterPatterns}(b-e)] evolve in comparison with the compact circular antinodes [Fig.~\ref{Fig2_FloaterPatterns}(a)] and grid-shaped node clusters [Fig.~\ref{Fig2_FloaterPatterns}(f)]. (c.1, c.2) The decaying part of $-M^*_2(\widetilde{r})$ is shown with the corresponding exponential fit. [Shaded gray rectangle in (c): Both the first minima and the second minima are marked. Remarkably, the second minima are only generated at intermediate $\phi$, neither at low $\phi$ nor high $\phi$.] The experimental values and the corresponding fits are presented by dots and solid lines, respectively. Note that, for (c.1), $\widetilde{r}=r-r_{\mbox{\tiny first minima}}$, and for (c.2), $\widetilde{r}=r-r_{\mbox{\tiny second minima}}$.}\label{Fig4_experimental_M0M1M2}
\end{figure*}

The calculated dimensionless functionals based on the experimental data are shown in Fig.~\ref{Fig4_experimental_M0M1M2} for values of $\phi$ spanning the whole range of patterns, i.e from antinode to node clusters. For all floater concentrations $\phi$, we find that when we increase $r$ the functionals decay due to merging of the disks: When $r$ becomes larger than the floater radius $R$, the discs start to overlap. Therefore, the normalized area of the resultant pattern $M^*_0(r)$ [Fig.~\ref{Fig4_experimental_M0M1M2}(a)] and the normalized perimeter $M^*_1(r)$ [Fig.~\ref{Fig4_experimental_M0M1M2}(b)] become smaller when $r>R$. The normalized Euler characteristic $M^*_2(r)$ [Fig.~\ref{Fig4_experimental_M0M1M2}(c)] first reaches to negative values, because when particles start to merge into clusters they first leave holes in between, each of which adds $-1$ to the Euler characteristic. Then, when $r$ increases further, the holes close and we are left with the number of clusters, i.e., $M^*_2(r)$ becomes positive again.

The corresponding experimental images that have been introduced in Fig.~\ref{Fig2_FloaterPatterns} can be instructive to understand the differences in the curves. The idea behind it is as follows: For compact and circular structures such as antinode clusters at low $\phi$ [Fig.~\ref{Fig2_FloaterPatterns}(a)], the functionals are expected to decay more quickly than for elongated antinode clusters [Fig.~\ref{Fig2_FloaterPatterns}(b)], heterogenous structures [Fig.~\ref{Fig2_FloaterPatterns}(c)], and filamentary structures [Fig.~\ref{Fig2_FloaterPatterns}(d)] at intermediate $\phi$. The reason is that the overlapping regions for compact structures are larger than those for filamentary structures at the same $r$ so that the functionals for compact structures decay sharply. This therefore suggests that the decay rate $\beta(\phi)$ [Fig.~\ref{Fig4_experimental_M0M1M2}(c)] can be used as a measure of the heterogeneity of the pattern: When we fit on exponential to the data, the decay is expected to be sharp for low and high $\phi$, where we have antinode and node clusters distributed over the wave surface in an organized way. Consequently, small $\beta(\phi)$ is expected. However, at intermediate $\phi$, as mentioned, elongated and filamentary structures occur in a more heterogeneous distribution relative to the antinode and the node clusters so that a smooth exponential decay occurs, i.e., large $\beta(\phi)$.

\begin{figure}[h!]
\begin{center}
  \includegraphics[width=8.0 cm]{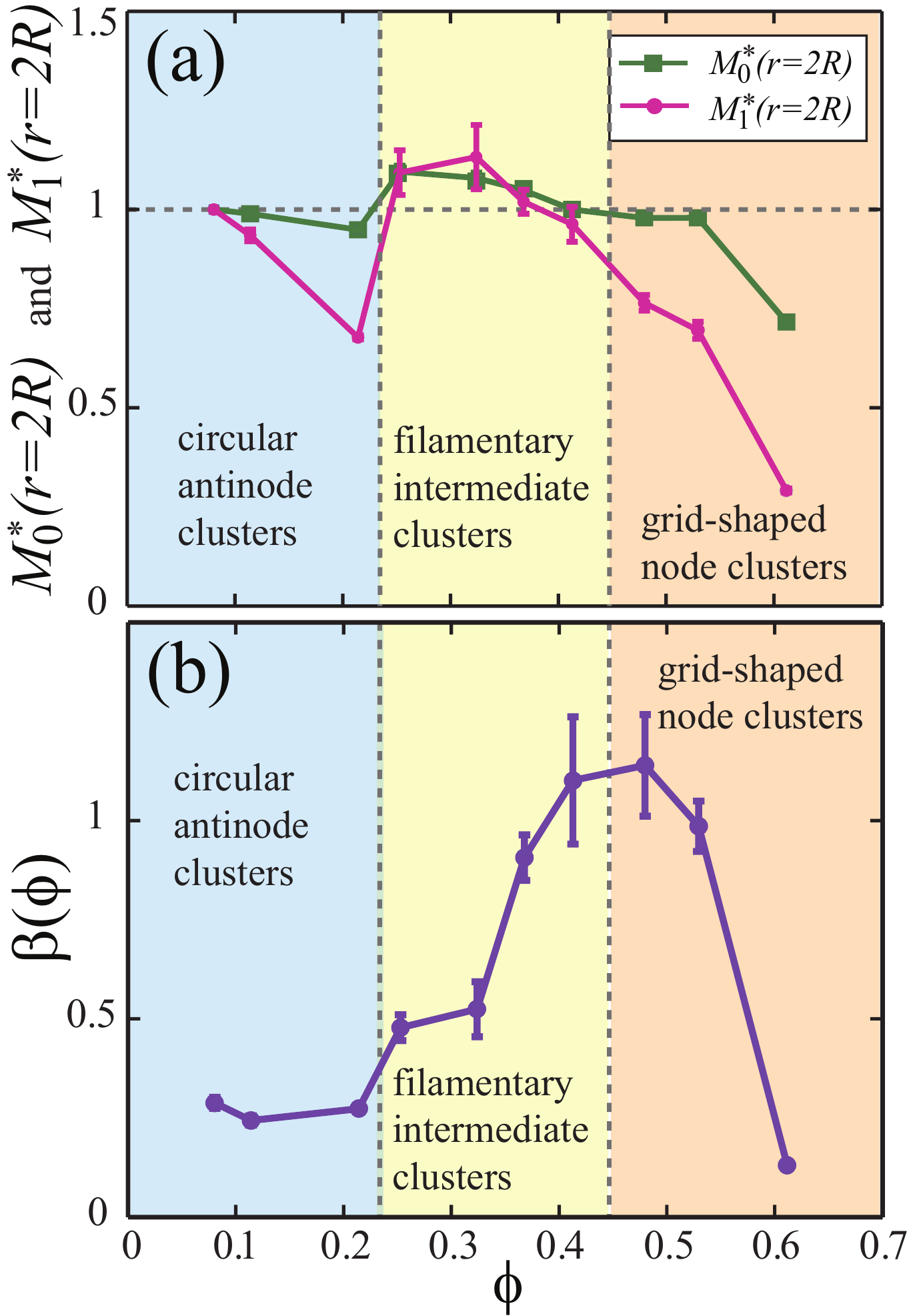}
\end{center}
\vspace{-6mm} \caption{\small Quantification of the decay of the three dimensionless Minkowski functionals. (a) The magnitude of the first two Minkowski functionals, namely $M^*_0(r)$ [area] and $M^*_1(r)$ [perimeter], for $r=2R$, normalized by those obtained at the lowest value of $\phi$. (b) The exponential decay rate $\beta(\phi)$ in the region where $-M^*_2(\widetilde{r})$ decays. All values increase for filamentary and more heterogeneous clusters at intermediate $\phi$, and they all tend to decrease when the clusters are more homogeneous and compact. (The mean values of each $\phi$ are the overall time averages of the experimental data. The error bars are the standard deviations of the time averages obtained by using a subset of a quarter of the data from the corresponding mean value.)}\label{Fig5_experimental_morphologyI}
\end{figure}

To investigate the validity of these arguments on the decay of the Minkowski functionals discussed here, we quantify the decay of $M^*_0(r)$, $M^*_1(r)$, and $M^*_2(r)$. Since $M^*_0(r)$ and $M^*_1(r)$ for all $\phi$ start to decay at the same point $r=R$, we quantify the decay by taking their value at $r=2R$. [For larger values of $r$ we obtain similar trends as a function of $\phi$ as presented in Fig.~\ref{Fig5_experimental_morphologyI}(a).] In Fig.~\ref{Fig5_experimental_morphologyI}(a), the magnitudes of $M^*_0(r=2R)$ and $M^*_1(r=2R)$ are presented as a function of $\phi$. Both are normalized by their values at the lowest $\phi$ so that they are equal to 1 for the lowest $\phi$, where we observe circular antinode clusters as demonstrated in Fig.~\ref{Fig2_FloaterPatterns}(a).

The situation is different for $M^*_2(r)$ [Fig.~\ref{Fig4_experimental_M0M1M2}(c)]. $M^*_2(r)$ first quickly descends to a negative minimum and then increases again towards a value around zero. Therefore, in this case we choose a different procedure to examine the corresponding decay: We look beyond the minimum (or minima, since for intermediate $\phi$ there are sometimes two minima before the decay sets in.) of $M^*_2(r)$ and try to locally fit an exponential $-M^*_2(r)\sim\exp(-r/\beta(\phi))$, as shown in Fig.~\ref{Fig4_experimental_M0M1M2}(c.1, c.2) in the gray shaded region of Fig.~\ref{Fig4_experimental_M0M1M2}(c). [The figure is split into two parts (c.1, c.2) for better visualization.] At low $\phi$ and the highest $\phi$, there is one minimum, however, at intermediate $\phi$, two minima exist. We apply the exponential fit to the region beyond the second minimum at intermediate $\phi$. $\beta(\phi)$ is the corresponding exponential decay rate. In Fig.~\ref{Fig5_experimental_morphologyI}(b), $\beta(\phi)$ is plotted as a function of $\phi$.

All decays indicate three regions: Circular (compact) antinode clusters at low $\phi$ where we observe sharp decays, i.e., small $M^*_0(r=2R)$, $M^*_1(r=2R)$, and $\beta(\phi)$, filamentary (heterogeneous) clusters at intermediate $\phi$ where we observe smooth decays, i.e., large $M^*_0(r=2R)$, $M^*_1(r=2R)$, and $\beta(\phi)$, and the decays become sharp again for grid-shaped node clusters at high $\phi$.

\subsection{Physical interpretation of the Minkowski functionals}\label{Sc:Physical interpretation of the Minkowski functionals}

We have shown that characterizing the decay of the functionals works to identify the morphological differences in the clustering patterns. However, a direct physical interpretation of the three Minkowski functionals is hard to provide. It is however feasible to construct quantities from the Minkowski functionals that do have an intuitive physical meaning. For example, if we divide the total area $[M_0(D)]_{\mbox{\footnotesize clusters}}$, which disregards the open spaces that are contained within the clusters as defined in the previous Subsection, by the number of clusters $[M_2(D)]_{\mbox{\footnotesize clusters}}$, we obtain the average cluster area $\overline{A}(r)$. We can do a similar division for the average perimeter $\overline{P}(r)$, leading to 
\begin{eqnarray}\label{Eq:clusterareaperimeteraspectratio}
\overline{A}(r)=\frac{[M_0(r)]_{\mbox{\footnotesize clusters}}}{[M_2(r)]_{\mbox{\footnotesize clusters}}},\,
\overline{P}(r)=\frac{[M_1(r)]_{\mbox{\footnotesize clusters}}}{[M_2(r)]_{\mbox{\footnotesize clusters}}}.
\end{eqnarray}

Finally, we define the average aspect ratio $\overline{S}(r)$ as $\overline{S}(r)\equiv\large(\overline{P}(r)\large)^2/4\pi\,\overline{A}(r)$ which gives

\begin{equation}\label{Eq:clusterareaperimeteraspectratio_IIPART}
   \overline{S}(r)=\frac{[M^2_1(r)]_{\mbox{\footnotesize clusters}}}{4\pi\,[M_0(r)]_{\mbox{\footnotesize clusters}}\,[M_2(r)]_{\mbox{\footnotesize clusters}}},
\end{equation}
where the dimensional Minkowski functionals $M_\nu(r)$ have been given in Eq.~\ref{Eq:2DPhysicalMinkowskifunctionals} and the decomposition described in Eq.~\ref{Eq:2DPhysicalMinkowskifunctionalsconnecteddisconnectedgroups} has been used~\cite{footnote1}. Note that $\overline{S}(r)$ has been defined such that it is $1$ for a set of $N$ disconnected circular regions of equal size.

\begin{figure*}[ht!]
\begin{center}
  \includegraphics[width=14.0 cm]{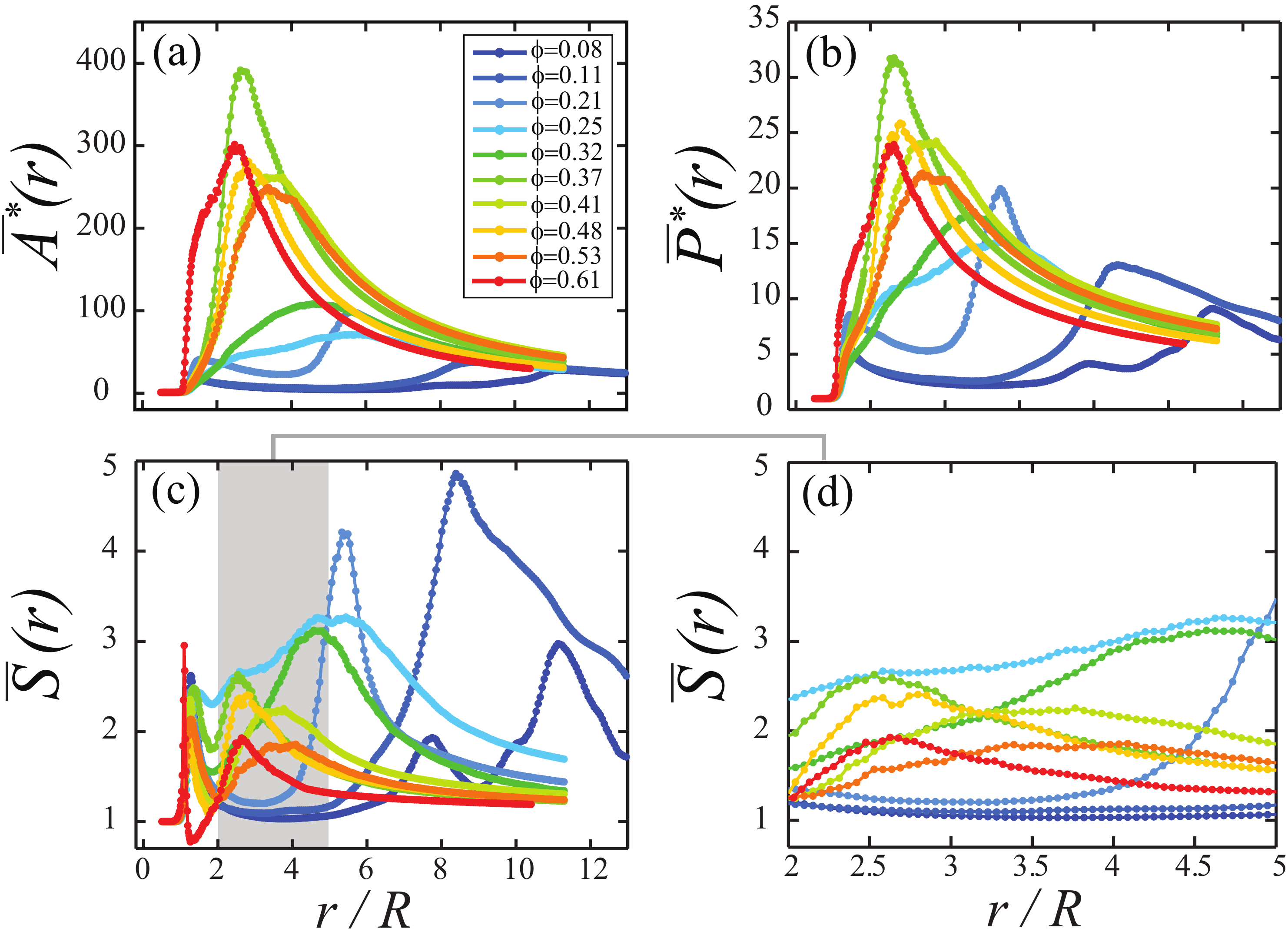}
\end{center}
\vspace{-6mm} \caption{\small Morphological characterization of the clustering patterns from antinode clusters (low $\phi$) to node clusters (high $\phi$) as a function of $r$ normalized by the floater radius $R$. (a) Average cluster area $\overline{A}^*(r)$, (b) average cluster perimeter $\overline{P}^*(r)$, and (c) average aspect ratio $\overline{S}(r)$. (d) The shaded region of the average aspect ratio plot (c) is enlarged. The interpretation of these plots is discussed in the main text.}\label{Fig6_experimental_Characterization}
\end{figure*}

Instead of $\overline{A}(r)$ and $\overline{P}(r)$, which for large $r$ are dominated by the area ($\pi r^2$) and perimeter ($2\pi r$) of a single disc, we use dimensionless $\overline{A}^*(r)$ and $\overline{P}^*(r)$ instead, which are defined similarly to the dimensionless Minkowski functionals [given in Eq.~\ref{Eq:2DPhysicalMinkowskifunctionalsdimensionlessforms}]
\begin{eqnarray}\label{Eq:clusterareaperimeteraspectratio}
\overline{A}^*(r)=\frac{[M^*_0(r)]_{\mbox{\footnotesize clusters}}}{[M^*_2(r)]_{\mbox{\footnotesize clusters}}}=\frac{\overline{A}(r)}{\pi r^2}, \\ \nonumber
 \overline{P}^*(r)=\frac{[M^*_1(r)]_{\mbox{\footnotesize clusters}}}{[M^*_2(r)]_{\mbox{\footnotesize clusters}}}=\frac{\overline{P}(r)}{2\pi r}.
\end{eqnarray}
When we calculate these quantities from the experimental data of our clustering patterns we obtain Fig.~\ref{Fig6_experimental_Characterization}.

Let us first look at the limits of these functionals, namely at $r\ll R$ and $r\gg R$. When $r\ll R$, none of the particles (disc) touch each other so that $N$ disconnected particles are expected. Then, the functionals become
\begin{eqnarray}\label{Eq:limitsrsmallerthanR}
  \overline{A}^*(r)\big|_{r\ll R}&\approx &\frac{\frac{N\,\pi r^2}{N\,\pi r^2}}{\frac{N}{N}}=1, \\ \nonumber
  \overline{P}^*(r)\big|_{r\ll R}&\approx &\frac{\frac{N\,2\pi r}{N\,2\pi r}}{\frac{N}{N}}=1, \\ \nonumber
  \overline{S}(r)\big|_{r\ll R}&\approx &\frac{N^2\,(2\pi r)^2}{4\pi\,N\pi r^2\,N}=1,
\end{eqnarray}
where $N$ is the number of floaters. In Fig.~\ref{Fig6_experimental_Characterization}, all experimental data satisfy this limit when $r\ll R$, i.e, all are equal to one. Secondly, let us consider the other limit, $r\gg R$. There, all disconnected groups are expected to merge, and to become a single cluster. Moreover, if $r$ becomes substantially larger than the distances between the individual particles, this final cluster will converge to a disc of radius $r$~\cite{footnote2} and the functionals become
\begin{eqnarray}\label{Eq:limitsrlargerthanR}
 \overline{A}^*(r)\big|_{r\gg R}&\approx &\frac{\frac{\pi r^2}{N\,\pi r^2}}{\frac{1}{N}}=1, \\ \nonumber
 \overline{P}^*(r)\big|_{r\gg R}&\approx &\frac{\frac{2\pi r}{N\,2\pi r}}{\frac{1}{N}}=1, \\ \nonumber
 \overline{S}(r)\big|_{r\gg R}&\approx &\frac{(2\pi r)^2}{4\pi^2\,r^2}=1.
\end{eqnarray}
Similar to the previous case, all experimental data is indeed seen to converge towards one when $r\gg R$ so that the limit is satisfied [see Fig.~\ref{Fig6_experimental_Characterization}].

Between these two distinct limits, all functionals have several maxima at different $r/R$. The first peaks in cluster area [Fig.~\ref{Fig6_experimental_Characterization}(a)] and cluster perimeter [Fig.~\ref{Fig6_experimental_Characterization}(b)] correspond to the event that individual particles merge into clusters, at a radius slightly larger than $R$, depending on how packed the structures are. The heights of the peaks, $\overline{A}^*_{\mbox{\footnotesize max}}$ and $\overline{P}^*_{\mbox{\footnotesize max}}$, respectively, give the typical size of these clusters. The corresponding $r_{\mbox{\footnotesize max}}$ where the peaks occur provides the average distance between the particles in the cluster, and therefore constitute a measure for the size of the porous region in between the particles of the cluster. For the three lowest values of $\phi$, we observe secondary peaks at a higher value of $r$. These peaks correspond to individual antinode clusters merging into larger structures that extend on the scale of the wavelength of the standing Faraday wave.

In Fig.~\ref{Fig6_experimental_Characterization}(c), we plot the third quantity we derived from the Minkowski functionals, namely the average aspect ratio $\overline{S}(r)$. The very first peaks, located at $r\approx R$, presumably correspond to touching particles merging into clusters of two or three, which generates large aspect ratios and should be disregarded. The location of the first peaks we find for $r>R$ [grey shaded area in Fig.~\ref{Fig6_experimental_Characterization}(c)] corresponds reasonably well to the ones in $\overline{A}^*(r)$ and $\overline{P}^*(r)$. The corresponding heights $\overline{S}_{\mbox{\footnotesize max}}$ provide the average aspect ratio of the main clusters in the pattern.

In Fig.~\ref{Fig6_experimental_Characterization}(d), the grey shaded area of Fig.~\ref{Fig6_experimental_Characterization}(c) is magnified to show the aspect ratio in this area of interest in greater detail: For the three lowest values of $\phi$, corresponding to the antinode clusters, we hardly find a hint of a maximum and the average aspect ratio is approximately equal to one. This stands to reason, since these antinode clusters are circular in shape and therefore hardly lead to any increase of the aspect ratio. That is to say, until above a certain value of $r$ the antinode clusters themselves start to merge, an effect that is clearly visible in the $\phi=0.21$ curve in Fig.~\ref{Fig6_experimental_Characterization}(d). For intermediate $\phi$ there is a clear and high maximum in the average aspect ratio for small $r/R\approx 2-3$ that corresponds to the formation of filamentary clusters. Finally, for the highest values of $\phi$ the aspect ratio goes down again, which is connected to the formation of the grid-shaped node clusters.

\begin{figure}[ht!]
\begin{center}
  \includegraphics[width=8.0 cm]{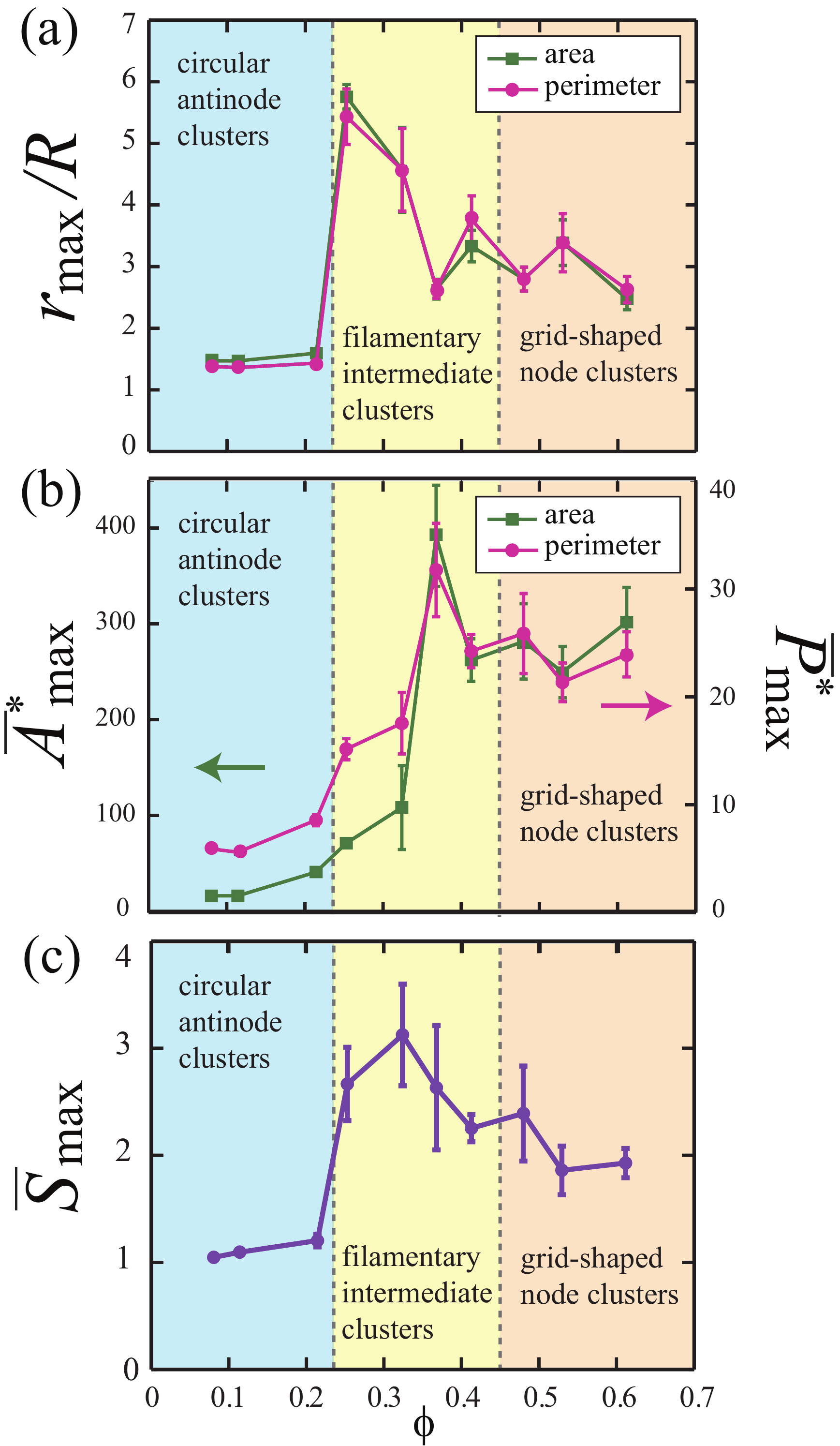}
\end{center}
\vspace{-6mm} \caption{\small Quantification of the morphological properties of the experimentally observed clustering patterns as a function of $\phi$, using the location and height of the maxima in Fig.~\ref{Fig6_experimental_Characterization}. (a) The location $r_{\mbox{\footnotesize max}}/R$ of the (first) maximum in the plots of $\overline{A}^*(r)$ [green squares] and $\overline{P}^*(r)$ [pink circles] versus $\phi$. (b) The height $\overline{A}^*_{\mbox{\footnotesize max}}$ [green squares] and $\overline{P}^*_{\mbox{\footnotesize max}}$ [pink circles] of the (first) maximum in the plots of $\overline{A}^*(r)$ (left vertical axis) and $\overline{P}^*(r)$ (right vertical axis) again versus $\phi$. (c) The height of $\overline{S}_{\mbox{\footnotesize max}}$ of the maximum in the gray shaded area of the plot of the aspect ratio $\overline{S}(r)$ versus $\phi$. The interpretation of these plots is discussed in the main text. (For each $\phi$, the mean values are the overall time averages of the experimental data. The error bars are the standard deviations of the time averages obtained by using a subset of a quarter of the data from the corresponding mean value.)}\label{Fig8_experimental_morphologyII}
\end{figure}

In Fig.~\ref{Fig8_experimental_morphologyII}(a), we plot the location $r_{\mbox{\footnotesize max}}$ of the first maximum of both $\overline{A}^*(r)$ and $\overline{P}^*(r)$ as a function of $\phi$. Since the two curves agree excellently with each other, we conclude that the average cluster area and perimeter reach their extreme values for the same $r$ for each $\phi$, which is plausible. Clearly, for low $\phi$, $r_{\mbox{\footnotesize max}}$ lies close to the particle radius $R$, indicating that neighboring particles in this antinode cluster regime are relatively close to one another in all spatial directions. At the boundary of the intermediate regime, the location of the maximum shifts quite suddenly to $r\approx6R$ as an indication of the formation of loose filamentary structures with holes in the order of 10 particle radii. When $\phi$ increases even further, the location of the maximum shifts back to lower values of $r$, because the system as a whole is becoming denser, and consequently the holes in between structures necessarily become smaller again. From the corresponding maxima themselves [Fig.~\ref{Fig8_experimental_morphologyII}(b)] a similar picture emerges. For the antinode clusters (small $\phi$), the average cluster size is small, and in the intermediate region rises quite steeply to an approximately constant large value, which it keeps inside the node cluster region (large $\phi$).

Finally, in Fig.~\ref{Fig8_experimental_morphologyII}(c), we plot the maximum value of the average aspect ratio $\overline{S}_{\mbox{\footnotesize max}}$ in the grey area of Fig.~\ref{Fig6_experimental_Characterization}(c). For low $\phi$, there is no maximum value and $\overline{S}(r)\approx 1$ throughout the shaded region, but then a finite maximum $\overline{S}_{\mbox{\footnotesize max}} \approx 3$ suddenly appears for $\phi = 0.25$, as evidence of the formation of structures that are elongated rather than circular. For higher $\phi$, the maximum $\overline{S}_{\mbox{\footnotesize max}}$ gradually decreases again and becomes approximately equal to 2 in the grid-shaped node clusters.

\section{Conclusion}\label{conc} 

The pattern analysis performed in this paper gives us an extensive characterization of the experimentally observed clustering patterns as a function of $\phi$. In our previous study~\cite{antinodenodePRE, CSanliThesis}, the clustering patterns have also been examined by introducing the global correlation coefficient $c(\phi)$, which nicely separate the antinode (low $\phi$) and the node (high $\phi$) clusters as a function of $\phi$ with a broad intermediate transition region. In agreement with $c(\phi)$, the intermediate transition region suggested by the global analysis lies in the same region as the one indicated by $c(\phi)$. (The analysis here suggests a slightly larger intermediate transition region.) In addition to what we could learn from $c(\phi)$, we can now also understand the detailed morphology of the clusters. Even though the Minkowski point pattern analysis developed in this paper mainly aims to quantifying the clustering patterns globally, local information can also deduced from the derived quantities such as $r_{\mbox{\footnotesize max}}$ as was discussed above [Fig.~\ref{Fig8_experimental_morphologyII}(a)]. 

We have characterized the morphology of the experimentally observed clusters as a function of the floater concentration $\phi$ from the antinode clusters to the node cluster regime. In a global picture, cluster area, perimeter, and aspect ratio are quantified utilizing the Minkowski functional point pattern approach. The results show the large variety of floater patterns generated as a result of the competition between the attractive capillary force and the drift due to the standing wave. On this large scale, we can distinguish three regions: circular antinode clusters (low $\phi$), filamentary and more heterogenous clusters (intermediate $\phi$), and grid-shaped node clusters (high $\phi$). This richness in the clustering patterns explains the broad transitional region between the antinode and node clusters, and why it is difficult to pinpoint the transition at a single $\phi$. In the energy argument of Refs.~\cite{antinodenodePRE, CSanliThesis}, only hexagonally symmetric patterns were designed without considering the rich morphological details of the structures that form at intermediate $\phi$. However, in this paper, we successfully characterize these details. In future work, one could improve the energy argument of the previous work by incorporating these details.

In conclusion, this paper is devoted to developing a detailed methodology to characterize the complex floater patterns we observed in the experiment. We believe that the methods developed here are applicable to other physical, chemical, biological, and also social systems that exhibit pattern formation, in order to learn more about common large scale motion observed in particulate flow, and so to examine to what degree there is universality in these large scale behaviors.

We thank S. Hilgenfeldt, S. Henkes, L. Tabourier, and R. Lambiotte for fruitful discussions. The work is part of the research program of FOM, which is financially supported by NWO. 


\begin{thebibliography}{apsrev}
\bibitem{footnote1} Mathematically, when moving from the strict (dimensional or dimensionless) Minkowski functionals $M_\nu(r)$ [Eq.~\ref{Eq:2DPhysicalMinkowskifunctionals}] and $M^*_\nu(r)$ [Eq.~\ref{Eq:2DPhysicalMinkowskifunctionalsdimensionlessforms}] to the $[M_0(r)]_{\mbox{\tiny clusters}}$, $[M^*_0(r)]_{\mbox{\tiny clusters}}$, $\overline{A}^*(r)$, $\overline{P}^*(r)$, and $\overline{S}(r)$ we loose an important property, namely that of additivity: If we calculate the Minkowski functionals of the union of two arbitrary domains $A\cup B$, the Minkowski functionals can be decomposed as $M_\nu(A\cup B)=M_\nu(A)+M_\nu(B)-M_\nu(A\cap B)$. The gain is however that the latter introduced quantities have a more direct physical interpretation.
\bibitem{footnote2} Note that we are evaluating the Minkowski functionals in an unbounded two-dimensional space, i.e., without boundaries (see Fig.~\ref{Fig3_MinkowskiPointPattern}). We allow the spheres to grow indefinitely in 2D space.
\end{thebibliography}
\end{document}